\documentclass[
    amsmath,
    amssymb,
    reprint,
    aps,
    prx,
    superscriptaddress,
    longbibliography
]{revtex4-1}

\usepackage[utf8]{inputenc}
\usepackage[T1]{fontenc}
\usepackage{physics}
\usepackage{nicefrac}
\usepackage{bbm, bm}
\usepackage{makecell}
\usepackage{multirow}
\usepackage{booktabs}
\usepackage{overpic}
\usepackage{tikz}
\usepackage{pgfplots}
\pgfplotsset{compat=1.18}
\usepackage[dvipsnames]{xcolor}
\usepackage{ragged2e}
\usepackage{bbm}

\usepackage[breaklinks=true, colorlinks]{hyperref}
\hypersetup{linkcolor=blue, citecolor=blue, urlcolor=MidnightBlue}
\usepackage{cleveref}

\begin{document}

\title{Whither quantum advantage? New classical 
lower bounds for Max-Cut on large-girth 3-regular 
graphs}

\author{Flaviano Morone}
\affiliation{Center for Quantum Phenomena, Department of Physics, New York University, New York, NY 10003 USA}

\author{Shivang Rawat}
\affiliation{Center for Quantum Phenomena, Department of Physics, New York University, New York, NY 10003 USA}

\author{Dries Sels}
\affiliation{Department of Physics, Boston University, 590 Commonwealth Ave., Boston, Massachusetts 02215, USA} 
\affiliation{Center for Computational Quantum Physics, Flatiron Institute, New York, NY, USA}

\begin{abstract}
Every 3-regular graph of girth at least $g$ 
has a cut containing at least a fraction 
$M_g$ of its edges, and the best rigorous 
bounds at large girth are 
$0.912\le \lim_{g\to\infty}M_g \le 0.92386$. 
Lower bounds on $M_g$ follow from the 
performance of local algorithms: on a graph 
of girth $g \ge 2p+2$ a depth-$p$ local 
algorithm sees the same tree around 
every edge, so its expected cut fraction on 
that tree bounds $M_g$ from below. 
We apply this strategy to the Variational 
Iterative Rotation Algorithm (VIRAL), a 
classical local algorithm, and evaluate it 
on the depth-$p$ tree for every even depth 
up to $p=32$. The resulting bounds exceed 
those recently obtained  from the Quantum 
Approximate Optimization Algorithm (QAOA) for every 
$p\ge 4$, by $0.005$ to $0.011$, and they extend 
the range of girths covered from $g=36$ to $g=66$. 
Notably, at $g=66$ we obtain $M_{66} \ge 0.9127$ 
at $99.99999\%$ confidence level, which raises 
the asymptotic lower bound from $0.912$ to $0.9127$. 
\end{abstract}

\maketitle

\begin{figure}[h!]
	\centering
        \vspace{-2mm}
	\includegraphics[width=0.5\textwidth]{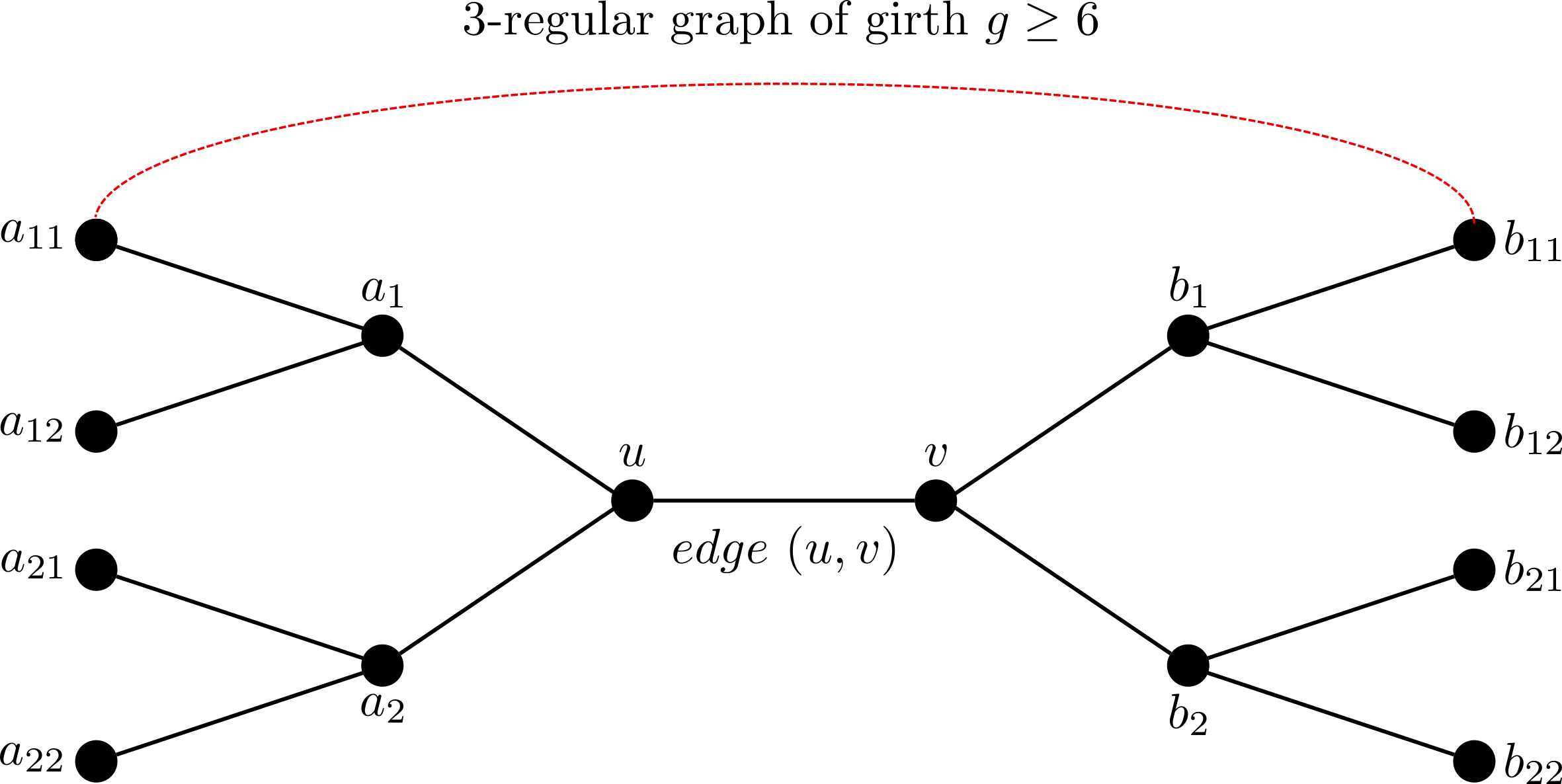}
	\caption{
    \justifying
    \textbf{Section of a 3-regular graph of girth 
    $g\geq6$}. The radius-$2$ neighborhood of an 
    edge $(u,v)$, consisting of the $14$ vertices within 
    distance $2$ of $u$ or $v$, and the $13$ edges incident 
    to $u$, $v$ and their neighbors. The girth forces 
    this tree structure. Any further edge among these 
    vertices must join two leaves (since every other 
    vertex already has degree $3$) and it is allowed only if the 
    two leaves lie on opposite sides of $(u,v)$, where it 
    closes a cycle of length $6$ (dashed edge). 
    }
    \label{fig:fig1}
    \vspace{-0mm}
\end{figure}

\emph{Introduction --}~The maximum cut of a 
graph $G=(V,E)$ is the largest number of edges 
connecting the two sides of a bipartition of 
$V$. Finding this cut is NP-hard even for 
3-regular graphs~\cite{yannakakis1978node}. 
The girth $g$ of a graph is the length of its 
shortest cycle. In a $3$-regular graph of girth 
$g$, the ball of radius $d$ around any vertex 
contains no cycles whenever $g\geq 2d+2$, and 
is then the same finite tree for every vertex, 
so all such graphs look alike locally and differ 
only globally, as illustrated in Fig.~\ref{fig:fig1} 
for $g\ge 6$. The question we address is what 
fraction of the edges an algorithm can guarantee 
to cut on these graphs. This motivates the following 
definition: 
\begin{equation}
M_g = \inf_{\substack{G \text{ 3-regular}\ 
\mathrm{girth}(G)\ge g}}\frac{\mathrm{MaxCut}(G)}{|E(G)|}\ .
\end{equation}
The quantity $M_g$ is non-decreasing in $g$ 
and $\lim_{g\to\infty}M_g$ exists, but its 
value is not known. The best lower bound, 
$\lim_{g\to\infty}M_g \ge 0.912$, follows by 
combining the results of Refs.~\cite{csoka2015invariant} 
and~\cite{gamarnik2018max}, as noted in~\cite{farhi2025lower}. 
The best upper bound comes from random 3-regular 
graphs, whose maximum cut fraction is asymptotically 
at most $0.92410$ by the 1RSB bound conjectured 
in~\cite{zdeborova2010conjecture} and proven 
in~\cite{coja2022ising}, and at most $0.92386$ by 
the tighter 2RSB bound of Ref.~\cite{harangi2025rsb}. Thus
\begin{equation}
0.912 \le \lim_{g\to\infty}M_g \le 0.92386\ ,
\label{eq:record}
\end{equation}
and any progress on the lower bound must be made 
inside a window of width $\Delta M \approx0.012$. 

\begin{figure}[ht]
	\centering
        \vspace{-2mm}
	\includegraphics[width=0.5\textwidth]{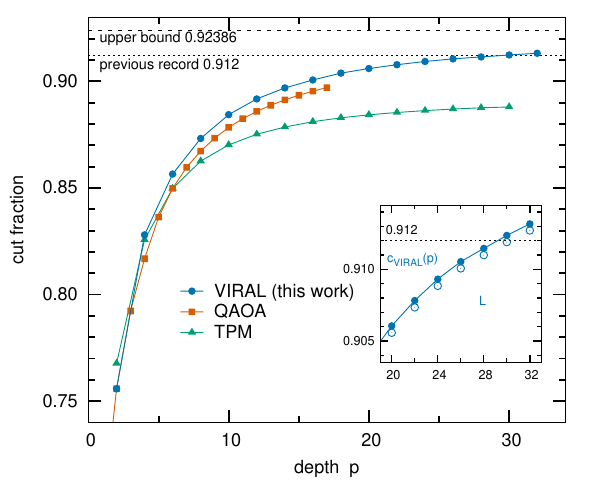}
	\caption{
    \justifying
    \textbf{Cut fraction versus depth on $3$-regular 
    graphs of girth $g\ge 2p+2$}. 
    Tree values $c_{\mathrm{VIRAL}}(p)$ of 
    Table~\ref{tab:viral-threshold} (circles), 
    QAOA values of Ref.~\cite{farhi2025lower} (squares) 
    and explicit vector algorithm (TPM) of 
    Ref.~\cite{thompson2021explicit} (triangles). 
    Statistical errors are smaller than the symbols. 
    The dashed line is the upper bound $0.92386$ 
    of Eq.~\eqref{eq:record}, the dotted line the 
    previous asymptotic lower bound $0.912$. 
    Inset: $p\ge 20$, with the lower confidence bound 
    $L$ of Table~\ref{tab:viral-threshold} (open circles), 
    which crosses $0.912$ at $p=32$. Lines are guides 
    to the eye. 
    }
    \label{fig:fig2}
    \vspace{-0mm}
\end{figure}

All lower bounds on $M_g$ to date come from 
local algorithms, which assign a vertex to 
a side of the cut using only information up 
to distance $p$ from that vertex. 
On a graph of girth $g\ge 2p+2$ the radius-$p$ 
neighborhood of every edge, defined below, is 
the same finite tree, as shown in Fig.~\ref{fig:fig1} 
for $g\geq6$ and $p=2$. Therefore, the expected 
cut fraction of a local algorithm of depth $p$ 
is a single number that can be computed on the tree 
and bounds $M_g$ from below for all $g\ge 2p+2$~\cite{farhi2025lower}. 
\begin{figure*}[ht]
	\centering
        \vspace{-2mm}
	\includegraphics[width=1\textwidth]{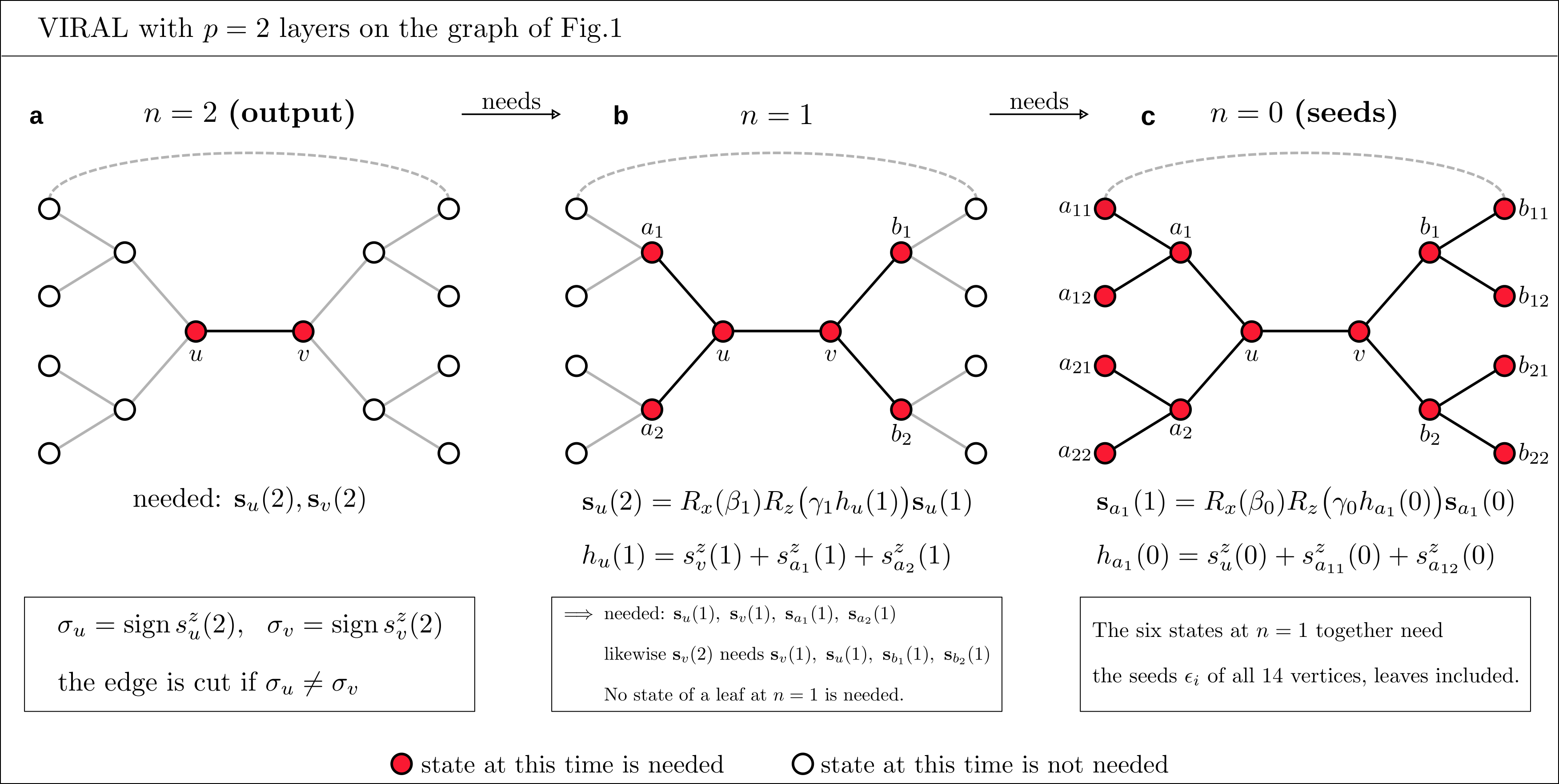}
	\caption{
    \textbf{VIRAL with $p=2$ layers on the graph of 
    Fig.~\ref{fig:fig1} traced backwards from the 
    output}. 
    {\bf (a)}, The cut variables $\sigma_u,\sigma_v$ 
    need ${\bf s}_u(2)$ and ${\bf s}_v(2)$. 
    {\bf (b)}, Through Eq.~\eqref{eq:viralmap} these 
    need the states at $n=1$ of $u$, $v$ and their 
    four other neighbors $a_1, a_2, b_1, b_2$.  
    No state of a leaf is needed at $n=1$ or $n=2$.
    {\bf (c)}, These six states need the seeds of all 
    $14$ vertices. The leaves enter only through their 
    seeds, so the edges joining two leaves, such as 
    the dashed one, play no role. Hence $(\sigma_u,\sigma_v)$ 
    is a fixed function of the 14 i.i.d. seeds on the 
    tree of Fig.~\ref{fig:fig1}, the same for every edge 
    of every 3-regular graph of girth $g\geq 6$.
    }
    \label{fig:fig3}
    \vspace{-0mm}
\end{figure*}
The Quantum Approximate Optimization Algorithm (QAOA)~\cite{farhi2014quantum} is an algorithm 
of this type and the depth-$p$ QAOA circuit is a 
local algorithm of radius $p$. A series of works 
evaluated the QAOA on large-girth regular graphs 
at increasing depth~\cite{wurtz2021fixed, basso2021quantum}, 
culminating in $M_g\ge 0.8971$ for $g\ge 36$ 
from $p=17$~\cite{farhi2025lower}. 
Classical local algorithms have been compared 
with the QAOA. At depth one~\cite{hastings2019classical} 
and two~\cite{marwaha2021local} simple classical 
rules outperform the QAOA for every degree, and 
numerical experiments on small random 3-regular 
graphs suggest the same at larger depth~\cite{barak2021classical}. 
On large-girth graphs, however, the picture 
is different: at degree three the explicit 
vector algorithm of Thompson, Parekh, and Marwaha~\cite{thompson2021explicit}, 
whose guarantees are known at every depth, 
falls behind the QAOA for $p\ge 7$~\cite{farhi2025lower} 
(see Table~\ref{tab:viral-threshold}). 
We recently introduced the Variational Iterative 
Rotation Algorithm (VIRAL)~\cite{morone2026variational}, 
a classical local algorithm in which every 
vertex carries a unit vector that is rotated 
$p$ times by angles set by its neighbors. 
On random 3-regular graphs with $10^6$ vertices, 
we numerically found that VIRAL cuts more edges 
than the QAOA at every depth $p\geq 4$ and its expected 
cut fraction extrapolates to $0.9231(5)$ for 
$p\to\infty$, very close to the upper bound in 
Eq.~\eqref{eq:record}~\cite{morone2026variational}. 
That benchmark, however, compares VIRAL on finite 
random graphs against the QAOA on the tree, and 
establishes no bound. Indeed, a random $3$-regular 
graph has short cycles, which at $N=10^6$ enter the 
radius-$p$ neighborhood of a typical edge already 
for $p\gtrsim9$, so a cut fraction measured there 
is not a bound on $M_g$. 

Here we evaluate VIRAL on the tree itself, which 
makes the comparison with the QAOA well defined 
and turns its cut fractions into bounds on $M_g$. 
We estimate the expected cut fraction on the depth-$p$ 
tree by sampling the seeds, with $10^7$ samples 
for every even $p\le 32$. The deepest computation 
involves $4\cdot 2^{32}\approx 1.7\times10^{10}$ 
vertices per sample. The resulting lower bounds on 
$M_g$ are reported in Table~\ref{tab:viral-threshold} 
and plotted in Fig.~\ref{fig:fig2}. 
They exceed the QAOA bounds and those of 
Ref.~\cite{thompson2021explicit} at equal depth 
for every $p\ge 4$, and they extend the range of 
girths covered from $g=36$ to $g=66$. At $g=66$ 
the estimate $c_{\mathrm{VIRAL}}(32)=0.913183(89)$ 
lies $13$ standard errors above the asymptotic 
record in Eq.~\eqref{eq:record}, and the rigorous 
bound is $M_{66}\ge 0.91271$ at $99.99999\%$ confidence 
level.

\emph{Setup --}~VIRAL assigns to every vertex $i$ a 
classical spin, i.e., a unit vector 
$\mathbf{s}_i=(s_i^x,s_i^y,s_i^z)\in S^2$, and 
updates all spins synchronously for $p$ layers 
as 
\begin{equation}
\begin{aligned}
\mathbf{s}_i(n+1) &= R_x(\beta_n)R_z\big(\gamma_n h_i(n)\big)
\mathbf{s}_i(n)\ ,\ \ \ n=0,\dots, p-1\ ,
\\
h_i(n) &= \sum_{j\in\partial i}s_j^z(n)\ ,
\end{aligned}
\label{eq:viralmap}
\end{equation}
where $\partial i$ is the set of neighbors of $i$, 
$R_z(\theta)$ and $R_x(\theta)$ are rotations by 
angle $\theta$ about the $z$ and $x$ axes, and 
$\beta_n,\gamma_n$ are fixed real parameters~\cite{morone2026variational}. The 
initial condition is 
\begin{equation}
\mathbf{s}_i(0) = \big(\cos\alpha,0,\epsilon_i\sin\alpha\big)\ ,
\label{eq:initcondition}
\end{equation}
with a fixed angle $\alpha$ and independent fair 
random seeds $\epsilon_i\in\{-1,+1\}$. After the 
last layer each spin is thresholded, 
\begin{equation}
\sigma_i(p)={\rm sign}(s_i^z(p))\ ,   
\end{equation}
and the cut is the set of edges with $\sigma_u\neq\sigma_v$. 
The $2p+1$ parameters 
$(\alpha,\beta_0,\dots,\beta_{p-1},\gamma_0,\dots,\gamma_{p-1})$ 
are trained once, as described below, and then 
frozen, so that the same parameters are used on 
every graph. The seeds are essential. On a regular 
graph, spins that start identical remain identical 
under Eq.~\eqref{eq:viralmap}, because every vertex 
sees the same field, $h_i=h$ for every $i$, and no 
edge would ever be cut. The seeds break this symmetry 
and are the only source of randomness in the algorithm. 
The initial condition in, given by expression~\eqref{eq:initcondition}, 
has a semiclassical origin. In the truncated Wigner 
approximation of the QAOA, the unitaries become the 
classical rotations of Eq.~\eqref{eq:viralmap} and 
the initial state $|+\rangle$ becomes a set of 
classical spins with (Gaussian) random transverse 
components, whose width plays the role of 
$\hbar$~\cite{sels2026absence}. The seeds $\epsilon_i$ 
are the minimal form of this noise: the transverse 
component has fixed amplitude $\sin\alpha$ and only 
its sign is random. In Ref.~\cite{sels2026absence} 
we found that the semiclassical dynamics with a tuned 
noise level reaches lower energies than the exact QAOA 
at every depth, and that replacing the Gaussian 
noise by the seeds and reoptimizing $\alpha$ 
together with the angles, which is VIRAL, improves 
the convergence of the energy to the ground-state 
value from $\log (p)/p$ to $1/p$. The optimal noise 
level found there vanishes as $p\to\infty$. In accordance, the trained amplitude for MaxCut also decreases with depth, from 
$\alpha=0.46$ at $p=2$ to $\alpha=0.018$ at $p=32$. 

Next we show that VIRAL is local, in the sense that 
whether an edge is cut depends only on a bounded 
portion of the graph around it. The graph enters 
Eq.~\eqref{eq:viralmap} only through the neighbor 
sets $\partial i$. Consider the graph of Fig.~\ref{fig:fig1} 
with $p=2$ and let us trace the computation of 
$\sigma_u=\mathrm{sign}(s^z_u(2))$ backwards, as 
in Fig.~\ref{fig:fig3}. By Eq.~\eqref{eq:viralmap}, 
$\mathbf{s}_u(2)$ is computed from $\mathbf{s}_u(1)$ 
and from the field 
$h_u(1)=s^z_v(1) + s^z_{a_1}(1) + s^z_{a_2}(1)$, so 
it needs the spins at time $1$ of $u$ and of its 
three neighbors, and the only edges involved are the 
three edges incident to $u$. 
Each of these four spins at time $1$ is in turn 
computed from the seed of its own vertex and from the 
seeds of the three neighbors of that vertex. Hence 
$\mathbf{s}_u(2)$ is a function of the seeds of the 
ten vertices within distance $2$ of $u$, and the only 
edges involved are those incident to $u,v,a_1,a_2$ 
(the vertices within distance $1$ of $u$). 
The leaves $a_{11},...,b_{22}$ enter only through 
their seeds and their states at times $1$ and $2$ 
are never needed, so which other vertices a leaf is 
connected to does not matter. In particular, the 
dashed edge of Fig.~\ref{fig:fig1} plays no role. 
Repeating the argument for $\sigma_v$, the pair 
$(\sigma_u, \sigma_v)$ is a function of the seeds 
of the 14 vertices within distance $2$ of $u$ or 
$v$, computed using only the solid edges of 
Fig.~\ref{fig:fig1}. 

The same unfolding applies at any depth $p$ and on 
any $3$-regular graph $G$, since each layer brings 
in the seeds of vertices one step farther away and 
the edges involved are those incident to the vertices 
whose state at some time $n\ge 1$ is needed. Thus, 
the pair $(\sigma_u, \sigma_v)$ at an edge $(u,v)$ 
is a function of the seeds of the vertices within 
distance $p$ of $u$ or $v$, computed using only the 
edges incident to the vertices within distance $p-1$ 
of $u$ or $v$. We call this set of vertices and edges 
the radius-$p$ neighborhood of the edge $(u,v)$. 
Nothing outside it can change whether the edge is 
cut or not. 

Now let $G$ have girth $g\ge 2p+2$. Then the 
radius-$p$ neighborhood of any edge contains 
no cycle, and is thus a tree. Since $G$ is 3-regular, 
it is always the same tree, consisting of the 
endpoints $u,v$ joined by the root edge, each 
carrying a complete binary tree of depth 
$p$, as in Fig.~\ref{fig:fig1} for $p=2$. 
Level $q$ of this tree is the set of vertices 
at distance $q$ from the nearer of $u$ and $v$, 
and contains $2^{q+1}$ vertices for $q=0,...,p$. 
In total the number of vertices in the tree is 
\begin{equation}
N_p = 2 + 4 + \dots +2^{p+1} = 4\cdot2^p-2\ .
\end{equation}
Since the algorithm sees nothing beyond this tree, 
and the seeds on it are i.i.d., the joint probability 
of $(\sigma_u,\sigma_v)$ is identical for every edge 
of every 3-regular graph of girth at least $2p+2$. 
We call 
\begin{equation}
c_{\mathrm{VIRAL}}(p) = 
\mathrm{Prob}\big[\sigma_u\neq \sigma_v\big]\ ,
\label{eq:cviral}
\end{equation}
where $(u,v)$ is the root edge of the depth-$p$ 
tree decorated with i.i.d. seeds, the value of 
the algorithm on the tree.
It is a property of the $2p+1$ parameters alone, 
not of any specific graph. 
Note that the tree itself is bipartite, so its own 
maximum cut fraction is $1$, attained by the 
alternating 2-coloring of its vertices. This coloring, 
however, is not a local function of the seeds. 
The reason is that the outputs of a local algorithm 
at two vertices at distance $\ell$ of the $3$-regular 
tree have a correlation that decays as $2^{\ell/2}$~\cite{backhausz2015ramanujan}, 
while the 2-coloring has correlation $(-1)^\ell$. 
Consistently, Eq.~\eqref{eq:newrecord} below 
shows that $c_{\mathrm{VIRAL}}(p)$ never exceeds 
$\lim_{g\to\infty}M_g \le 0.92386$.

\emph{From the tree to a bound on $M_g$ --}~Let us 
consider any 3-regular graph $G$ of girth 
$g\ge 2p+2$ and run the algorithm on it 
with the fixed parameters and fresh i.i.d. seeds 
$\epsilon$. The output $\sigma(\epsilon)$ is a random 
cut of $G$, whose expectation is 
\begin{equation}
\mathbb{E}\left[\frac{\mathrm{cut}(\sigma(\epsilon))}{|E|}\right]=
\frac{1}{|E|}\sum_{(u,v)\in E}
\mathrm{Prob}\big[\sigma_u\neq\sigma_v\big] = c_{\mathrm{VIRAL}}(p)\ .
\label{eq:cutviral}
\end{equation}
Now, a random variable is at least as large as its 
mean with positive probability and, since there are 
finitely many seed configurations, some $\epsilon^*$ 
satisfies 
$\mathrm{cut}(\sigma(\epsilon^*))/|E|\ge c_{\mathrm{VIRAL}}(p)$, and the maximum cut is at least as large, 
i.e.
\begin{equation}
\frac{\mathrm{MaxCut}(G)}{|E(G)|} \ge c_{\mathrm{VIRAL}}(p)\ .
\label{eq:viralbound}
\end{equation}
Note that nothing about $G$ was used except its degree 
and its girth, so Eq.~\eqref{eq:viralbound} holds for 
every graph in the class defining $M_g$, and thus for 
their infimum. Together with the fact that $M_g$ is 
non-decreasing in $g$ we conclude that 
\begin{equation}
\lim_{g\to\infty}M_{g}\ge M_{2p+2}
\ge c_{\mathrm{VIRAL}}(p)\ .
\label{eq:newrecord}
\end{equation}
This is precisely the argument by which Farhi 
\textit{et al.} turn a QAOA expectation value into 
a bound on $M_g$~\cite{farhi2025lower}. The 
only difference is that the expectation in 
Eq.~\eqref{eq:cutviral} is over classical random 
seeds rather than over a quantum state. 

Two consequences follow. First, the bound is constructive, 
in that on any 3-regular graph of girth at least $2p+2$, 
one run of the algorithm costs $O(p|E|)$ operations and 
produces a cut whose expected fraction is $c_{\mathrm{VIRAL}}(p)$, 
with the parameters precomputed once for each $p$.
Second, $c_{\mathrm{VIRAL}}(p)$ is a classical 
expectation that we estimate by sampling the seeds, 
whereas the QAOA expectation of Ref.~\cite{farhi2025lower} 
is evaluated exactly by tensor-network methods. 
Each sample returns one bit, i.e., whether the root 
edge is cut, so from $S$ samples with $K$ cut events 
the estimate $\hat{c}=K/S$ has standard error 
$\sigma= \sqrt{\hat{c}(1-\hat{c})/S}$. For $S=10^7$ 
the estimate is Gaussian to high accuracy, and the 
probability that it exceeds the true value by more 
than $5.2\sigma$ is $10^{-7}$. Hence 
\begin{equation}
L= \hat{c} - 5.2\sigma 
\label{eq:L}
\end{equation}
is a lower bound  on $c_{\mathrm{VIRAL}}(p)$ at 
$99.99999\%$ confidence level. 
Table~\ref{tab:viral-threshold} reports the exact 
binomial bound at the same level, which agrees 
with Eq.~\eqref{eq:L} to within $10^{-6}$. Since 
Eq.~\eqref{eq:newrecord} holds for every value of 
$c_{\mathrm{VIRAL}}(p)$, the statement $M_g\ge L$ 
for all $g\ge 2p+2$ holds at the same confidence level. 

\begin{table*}[t]
\begin{ruledtabular}
\begin{tabular}{rrllll}
$p$ & $g\ge$ & $c_{\mathrm{VIRAL}}(p)$ & $L_{99.99999\%}$ & QAOA & TPM \\
\hline
2  & 6  & 0.755779(136) & 0.75507 & 0.7559 & 0.7677 \\
4  & 10 & 0.827983(119) & 0.82736 & 0.8168 & 0.8257 \\
6  & 14 & 0.856546(111) & 0.85596 & 0.8498 & 0.8498 \\
8  & 18 & 0.873223(105) & 0.87267 & 0.8673 & 0.8626 \\
10 & 22 & 0.884378(101) & 0.88385 & 0.8784 & 0.8702 \\
12 & 26 & 0.891760(98)  & 0.89124 & 0.8859 & 0.8752 \\
14 & 30 & 0.896941(96)  & 0.89644 & 0.8913 & 0.8786 \\
16 & 34 & 0.900672(95)  & 0.90017 & 0.8954 & 0.8811 \\
18 & 38 & 0.903873(93)  & 0.90338 & 0.8971 & 0.8829 \\
20 & 42 & 0.906053(92)  & 0.90557 & ---    & 0.8843 \\
22 & 46 & 0.907825(91)  & 0.90734 & ---    & 0.8854 \\
24 & 50 & 0.909322(91)  & 0.90884 & ---    & 0.8863 \\
26 & 54 & 0.910545(90)  & 0.91007 & ---    & 0.8870 \\
28 & 58 & 0.911469(90)  & 0.91100 & ---    & 0.8876 \\
30 & 62 & 0.912373(89)  & 0.91190 & ---    & 0.8880 \\
32 & 66 & 0.913183(89)  & 0.91271 & ---    & ---    \\
\end{tabular}
\end{ruledtabular}
\caption{Tree value $c_{\mathrm{VIRAL}}(p)$ 
from $S=10^{7}$ samples per depth (statistical
error on the last digits in parentheses). The 
exact one-sided lower confidence bound $L$ on 
$c_{\mathrm{VIRAL}}(p)$ at $99.99999\%$ confidence 
level, truncated to five decimals, which 
bounds $M_g$ from below for all $g\ge 2p+2$,  
and the lower bounds obtained at the same depth 
from the QAOA~\cite{farhi2025lower} and from 
the explicit vector algorithm (TPM) of 
Ref.~\cite{thompson2021explicit}. The QAOA entry 
at $p=18$ is the deepest available value, $p=17$, 
as no QAOA values have been computed beyond that 
depth.}
\label{tab:viral-threshold}
\end{table*}

\emph{Computing $c_{\mathrm{VIRAL}}(p)$ --}~The depth-$p$ 
tree is not a truncation of the infinite tree but 
the exact object to consider, in that nothing beyond 
it reaches the root edge. We draw the $N_p$ i.i.d. 
seeds, run Eq.~\eqref{eq:viralmap} on the tree, and 
record the single bit $\mathbbm{1}[\sigma_u\neq\sigma_v]$. 
Then $c_{\mathrm{VIRAL}}(p)$ is the mean of this bit 
over $S=10^7$ independent draws. In particular, at 
$p=2$ no sampling is needed: the tree has $14$ vertices, 
as in Fig.~\ref{fig:fig1}, so all $2^{14}=16384$ seed 
configurations can be run through Eq.~\eqref{eq:viralmap} 
one by one, and we find that $12380$ of them cut the 
root edge. The exact value 
$c_{\mathrm{VIRAL}}(2) = 12380/16384=0.755615$ agrees 
with the sampled one in Table~\ref{tab:viral-threshold} 
within $1.2$ standard errors.

\emph{Parameters --}~The $2p+1$ parameters are 
those obtained in Ref.~\cite{morone2026variational} 
by training VIRAL on random $3$-regular graphs, 
and are used here unchanged \footnote{The parameters $(\alpha,\beta_n,\gamma_n)$ for every depth $p$ and the code to reproduce all results are available at \url{https://github.com/shivangrawat/viral-tree-maxcut}.}. We computed them 
for even $p$ only, which is why Table~\ref{tab:viral-threshold} 
lists even depths. Each training instance 
consists of a random $3$-regular graph with 
$N=10^6$ vertices and one fixed realization 
of the seeds. The cost function is the 
continuous energy $\sum_{(i,j)\in E} s^z_i(p)s^z_j(p)$ 
averaged over $10$ instances, whose gradient 
with respect  to $(\alpha,\beta_n,\gamma_n)$ 
is computed exactly by backpropagation through 
the $p$ layers of Eq.~\eqref{eq:viralmap}. 
Two remarks are in order: first, the same 
parameters are used on every graph, as the argument 
leading to Eq.~\eqref{eq:newrecord} requires; second, 
they are not optimal for the tree, since 
they were optimized on graphs with cycles. 
Therefore, every value in Table~\ref{tab:viral-threshold} 
is a conservative estimate of what VIRAL can 
guarantee at that depth, in that parameters 
optimized on the tree itself can only do better.

\emph{Discussion --}~Evaluated on the depth-$p$ 
tree, VIRAL yields rigorous lower bounds on $M_g$ 
for every girth up to $66$. At equal depth they 
exceed the QAOA bounds by $0.005$ to $0.011$ and 
those of Ref.~\cite{thompson2021explicit} at every 
$p\ge 4$, and at $g=66$ they pass $0.912$, a value 
previously reached  only in the limit of infinite 
girth. Therefore, the asymptotic lower bound in 
Eq.~\eqref{eq:record} becomes 
\begin{equation}
0.9127 \le \lim_{g\to\infty}M_g \le 0.92386\ .
\end{equation}
The bounds are constructive, and the algorithm 
attaining them runs in time linear in the number 
of edges. 

These results bear on the question of quantum advantage 
for local algorithms. Ref.~\cite{farhi2025lower} 
found that on $3$-regular graphs the QAOA beats 
the best classical local algorithm with a provable 
guarantee for $g\ge 16$, and Ref.~\cite{apte2026provable} 
has recently extended this comparison to Max-$k$-Cut 
on $d$-regular graphs, where the QAOA overtakes an 
improved vector algorithm for $g\ge 20$ at every $k$ 
and $d$ considered, an apparent quantum advantage at 
moderate girth. The tree value $c_{\mathrm{VIRAL}}(p)$ 
is a guarantee of the same kind, and at $k=2$ and $d=3$ 
it exceeds the QAOA value at every even depth $p\ge 4$ 
up to $32$, in particular at every girth where 
Ref.~\cite{farhi2025lower} found the QAOA ahead 
of the classical algorithms. 
In this case, therefore, the advantage does not exist. 
Whether the same holds at higher degree and for $k\ge 3$ 
is a natural next test. The same conclusion was 
reached on the Sherrington-Kirkpatrick (SK) model, 
which governs the large-degree limit of Max-Cut 
on regular graphs~\cite{basso2021quantum}. There 
VIRAL outperforms the QAOA at every depth~\cite{morone2026variational}, 
and so does the semiclassical limit of the QAOA 
itself~\cite{sels2026absence}. In both cases, the 
dense SK model and sparse $3$-regular graphs, the 
quantum fluctuations of the QAOA are not a resource. 
Their classical remnant, i.e. the random seeds, is 
all the dynamics needs, and it does better without 
the rest. 

Finally, we note that the increments 
$c_{\mathrm{VIRAL}}(p)-c_{\mathrm{VIRAL}}(p-2)$ are 
still decreasing at $p=32$ and the present data do not 
determine the limit $p\to\infty$. How much of the 
remaining gap to $0.92386$ a local algorithm can close 
is an open question.

\medskip

\emph{Acknowledgements}
The Flatiron Institute is a division of the 
Simons Foundation. D.S. and F.M were partially 
supported by AFOSR under Award No. FA9550-21-1-0236 
and by ONR under Award No. N00014-23-1-2771. This 
work was supported in part through the NYU IT High 
Performance Computing resources, services, and 
staff expertise. Correspondence should be addressed 
to F.M. at fm2452@nyu.edu

\newpage
\bibliography{references}

\end{document}